# Extraction de termes, reconnaissance et labellisation de relations dans un thésaurus – Vers une ontologie

Marie-Noelle BESSAGNET (1)
marie-noelle.bessagnet@univ-pau.fr

Eric KERGOSIEN (2)
eric.kergosien@univ-pau.fr

Mauro GAIO (2)
mauro.gaio@univ-pau.fr

*UPPA, Laboratoire LIUPPA, IAE, Avenue du doyen Poplawski, 64012 PAU (1)*

*UPPA, Laboratoire LIUPPA, Faculté des Sciences, Département Informatique, 64000 PAU (2)*

**Résumé** : Dans le domaine des systèmes de documentation, l'usage des thésaurus à des fins d'indexation puis de recherche d'information est courant voire obligatoire. Dans les bibliothèques et les médiathèques francophones, par exemple, les documents possèdent de par le travail effectué par les bibliothécaires de riches informations de description, sous la forme de notices descriptives, décrites sur la base du thésaurus RAMEAU. Nous exploitons ces deux types de ressources (documents et notices) afin de créer une première structure sémantique représentant le travail d'indexation des bibliothécaires pour élaborer le thésaurus TERRIDOC. Notre corpus de référence a une forte connotation territoriale. Nous nous intéressons également à la transformation de thésaurus en ontologie de domaine. En effet, nous souhaitons obtenir une ontologie de domaine offrant une représentation synthétique du territoire implicitement décrit par le fonds documentaire traité, en faisant appel à des ressources externes de type SIG.

**Abstract** : Within the documentary system domain, the integration of thesauri for indexing and retrieval information steps is usual. In libraries, documents own rich descriptive information made by librarians, under descriptive notice based on Rameau thesaurus. We exploit two kinds of information in order to create a first semantic structure. A step of conceptualization allows us to define the various modules used to automatically build the semantic structure of the indexation work. Our current work focuses on an approach that aims to define an ontology based on a thesaurus. We hope to integrate new knowledge characterizing the territory of our structure (adding "toponyms" and links between concepts) thanks to a geographic information system (GIS).

## 1 Introduction

Les bibliothèques et les médiathèques, renferment des corpus documentaires de type patrimonial conséquents de plus en plus facilement disponibles pour le grand public grâce au format électronique (numérisation et OCRisation). Cependant l'accessibilité par le grand public à ces corpus reste encore problématique. Dans ces organismes de conservation, chaque document est associé à une notice descriptive établie par les bibliothécaires, elles sont construites sur la base d'un thésaurus faisant autorité dans le milieu, le thésaurus RAMEAU[1]. Nous proposons une exploitation automatique de ces deux types de ressources afin de créer une structure sémantique représentant le travail d'indexation des bibliothécaires.

Nous souhaitons, d'une part, proposer aux bibliothécaires des outils de visualisation et de parcours de cette structure afin de valider leur travail d'indexation. Cette approche se décompose en deux phases : la première étant d'identifier et de représenter l'information à l'aide des connaissances expertes extraites automatiquement des notices, la deuxième étant de donner la possibilité de naviguer dans le fonds documentaire via les connaissances identifiées pour faciliter la représentation du travail d'indexation. Actuellement, les relectures et éventuelles corrections sont réalisées manuellement notice par notice, rendant cette tâche fastidieuse. Nous pensons que la représentation sous forme de carte de connaissances extraites automatiquement du travail d'indexation apporte un premier élément de réponse à leurs attentes en leur offrant une synthèse exhaustive d'un état de l'indexation de la base documentaire. Nous nous intéressons, d'autre part, à la conceptualisation d'un sous-ensemble du thésaurus RAMEAU afin de produire une représentation ontologique de domaine mettant en

---
[1] Répertoire d'autorité-matière encyclopédique et alphabétique unifié ; http://rameau.bnf.fr/. Thésaurus défini au sein de la Bibliothèque Nationale de France (BNF)

avant un territoire. RAMEAU a été adopté dans le contexte d'informatisation des bibliothèques françaises dans les années 80. Cette liste d'autorités s'inspire largement du langage d'indexation RVM Laval (Canada), qui lui-même est issu d'un long travail de traduction à partir des vedettes-matières américaines tirées des LCSH (Library of Congress Subject Headings). Les thésaurus sont des vocabulaires contrôlés de termes représentant généralement un domaine particulier gérant des relations hiérarchiques, associatives et d'équivalence. On peut citer NML's Medical Subject Headings (MeSH) dans le domaine médical pour indexer et rechercher des articles, le célèbre Wordnet, plus général, utilisé dans des travaux d'analyse sémantique. Dans le contexte de transformation de thésaurus en ontologie, le W3C travaille sur un méta schéma de référence, le SKOS (Simple Knowledge Organization System), basé sur les concepts [1].

Dans notre cas, l'objectif est d'expliciter la sémantique informelle du thésaurus autour des termes décrivant un territoire en se restreignant à l'aspect spatial. L'information du corpus (ici les notices descriptives) peut nous aider à spécifier des relations entre termes pouvant être ambigües dans un thésaurus afin de créer une première ontologie, L'analyse linguistique automatisée de ces notices doit ensuite nous permettre d'enrichir l'ontologie du domaine par de nouveaux concepts qualifiant un territoire.

Dans une première partie (&2), nous présenterons les problématiques et objectifs de notre travail de recherche. Nous développerons les travaux connexes dans le (&3) puis notre approche (&4) pour construire de manière automatique un thésaurus particulier : le thésaurus TERRIDOC. Enfin, nous expliciterons notre démarche pour passer d'un thésaurus particulier à une ontologie de domaine (&4) puis nous conclurons (&5).

## 2 Problématiques et objectifs

L'intérêt de disposer d'un fonds documentaire et de pouvoir ensuite proposer à des utilisateurs d'accéder aux informations nécessaires pour leur activité est primordial. Cela implique, d'une part, la nécessité d'identifier les informations pertinentes et d'autre part, la possibilité de fournir des moyens pour y accéder. Les possibilités pour organiser, classer et structurer un ensemble de documents sont nombreuses. Ainsi, afin d'offrir aux experts du domaine[2] un outil de validation de l'utilisation du langage contrôlé qu'ils ont mis en œuvre pour harmoniser leurs formulations de thèmes décrivant le contenu des documents, nous avons élaboré dans notre démarche deux phases préalables : (i) Extraction et Structuration des connaissances du domaine du fonds documentaire ; (ii) Navigation et interrogation du fonds documentaire en proposant une représentation sémantique de ce dernier. L'un de nos objectifs est la mise en place d'un processus pour passer d'un thésaurus classique à une base de connaissances. Ainsi, la première phase de notre démarche permet de créer automatiquement une structure représentant sous forme de thésaurus (le thésaurus TERRIDOC) le travail d'indexation des bibliothécaires en nous appuyant sur les notices descriptives pour identifier les termes et sur RAMEAU pour extraire les relations entre ces termes. Chacun des termes est ainsi enrichi par les relations de type « employé pour », « terme associé » et « terme générique » et par les termes RAMEAU se trouvant liés par ces relations. Ainsi, nous considérons chaque terme extrait des notices descriptives comme terme « de bas niveau » car rattaché directement à des documents et nous enrichissons le thésaurus avec les termes plus génériques du thésaurus RAMEAU. Le but visé par l'enrichissement du thésaurus via ces termes génériques est de permettre le regroupement en une seule structure des termes extraits. L'étape suivante consiste à enrichir cette première structure sémantique par des connaissances renseignant sur le territoire implicitement décrit par le fonds documentaire dans le but d'offrir aux utilisateurs un accès élargi à l'information. Ainsi, nous cherchons à exploiter dans nos ressources trois types d'informations : nous les qualifions d'entité thématique, d'entité spatiale (ES) [2] et d'entité temporelle (dans cet article, nous ne traiterons pas ce dernier type). Afin de capter ces entités et les relations existantes entre ces dernières, nous avons mis en place une chaîne de traitement sémantique automatisée, développée grâce à l'environnement Linguastream[3]. Elle est composée de quatre grandes phases [3]: (a) la lemmatisation pour segmenter les mots ; (b) l'analyse lexicale et morphologique pour la reconnaissance des mots ; (c) l'analyse syntaxique, basée sur des grammaires, afin de trouver les relations entre les mots ; (d) enfin l'analyse sémantique pour réaliser une interprétation plus spécifique sur les syntagmes retenus.

Afin de détecter ces entités, la partie extraction est découpée en étapes. La première (1) concerne la collecte d'ouvrages numérisés relatant d'un territoire. La seconde (2) supporte une analyse linguistique puis sémantique afin d'extraire les Entités précitées. La troisième (3) s'appuie d'une part sur des ressources géographiques (communes, lieux-dits, routes, pics, vallées, …) afin de valider les ES détectées à l'étape précédente et d'autre part sur la ressource RAMEAU afin de valider les Entités Thématiques. La dernière étape (4) propose la labellisation des relations entre ces diverses entités. Au vu de l'analyse de notre corpus, nous souhaitons nous intéresser à l'ensemble des relations binaires suivantes : Entité Thématique- Entité Spatiale et Entité

---

[2] Nous collaborons avec les bibliothécaires de la Médiathèque Intercommunale à Dimension Régionale (MIDR) de Pau)

[3] http://www.linguastream.org/whitepaper.html

Thématique- Entité Temporelle. Nous aborderons dans ce papier la relation Entité Thématique- Entité Spatiale. A cet effet, nous montrerons la démarche pour détecter des qualificatifs des toponymes ainsi que des relations d'approximation de sens avec les termes du thésaurus.

## 3   Travaux connexes

Transformer des thésaurus en ontologie fait l'objet de travaux de recherche récents. Depuis plusieurs années, les ontologies sont créées et utilisées dans le domaine de l'ingénierie des connaissances et notamment leur représentation. Le champ d'application est très large [4] : d'une manière générale dans l'indexation et la recherche d'information, et plus particulièrement dans le domaine médical, dans le domaine touristique, dans le domaine de l'éducation, dans le domaine de l'héritage culturel.

La conception automatique d'ontologies émerge comme un sous-domaine de l'ingénierie des connaissances. Afin de créer ces ontologies, il existe diverses approches et méthodes. Certains travaux reposent sur l'analyse de textes afin d'aider à la construction semi automatique des ontologies. D. Bourigault et al [5] décrivent les quatre étapes de la méthodologie de construction d'une ontologie à partir de textes (constitution du corpus à partir d'une analyse des besoins de l'application, étude linguistique afin d'identifier les termes et relations constituant la structure sémantique, normalisation sémantique définissant dans un langage formel les concepts et relations identifiées, validation de la formalisation par des spécialistes du domaine étudié). On peut remarquer que pour bâtir une ontologie à partir de textes, on utilise soit des ressources linguistiques externes, soit le corpus constitué des documents. Les outils supportant ces méthodes utilisent des techniques linguistiques pour retrouver les formes terminologiques dans l'analyse des textes. A Maedche et S. Staa [6] décrivent différents types d'approches distinguées en fonction du support sur lequel elles se basent : les plus courantes sont comme ci-dessus à partir de textes, de dictionnaires, d'autres à partir de bases de connaissances, ou encore de schémas semi-structurés et de schémas relationnels. Les travaux de [7] et [8] proposent une approche permettant de construire une ontologie minimale ; le processus consiste " *in extracting from texts specific types of information, rather than general-purpose relations. Accordingly, they produced remarkable efforts to conceptualize their competence domain through the definition of a core ontology*".

Comme déjà mentionné, nous nous intéressons plus particulièrement aux méthodes permettant de transformer un thésaurus en ontologie du domaine. Dans [9], l'approche présentée permet de transformer le thésaurus à facettes de l'art et de l'architecture AAT en ontologie pour indexer des images. Cette approche est entièrement manuelle. L'ontologie est formalisée en RDFS. Deux étapes d'identification de concepts et d'augmentation des concepts grâce à des propriétés permettent de définir cette ontologie. La méthode explicitée dans [10] repose sur trois étapes. Cette dernière a permis la transformation du thésaurus AGROVOC couvrant le domaine de l'agriculture, de la forêt, de la nourriture et des domaines reliés tel que l'environnement. L'originalité se base sur une phase d'apprentissage afin d'extraire des relations supplémentaires augmentant ainsi la sémantique liée au thésaurus de base. Nos travaux actuels se rapprochent de ceux développés d'une part par [11], [12] et [13] qui s'appuient sur un thésaurus et un langage ontologique tel OWL pour améliorer l'interopérabilité entre outils et pour donner accès à ce dernier à une plus large communauté et d'autre part ceux de [14] qui simplifient l'opération de création d'ontologie à travers une approche permettant d'enrichir un thésaurus pour créer une ontologie à partir de sources de connaissances du domaine (vocabulaires, thésaurus, etc). Ces sources formalisées, contenant des termes représentant le domaine et (pour les thésaurus) des relations entre ces termes, apportent alors un plus sémantique indéniable à la représentation du domaine étudié.

En accord avec [11], l'une des étapes importantes pour transformer un thésaurus en ontologie est d'avoir une représentation des concepts et de leurs relations dans un format « traitable » par une machine. Nous avons choisi, dans un premier temps, de formaliser notre structure sémantique du domaine sur la base des Topics Map et sur OWL. D'une part, les TM sont le formalisme le plus adapté à des fins de navigation dans la carte de concepts[4] et dans leurs instances, ce qui nous a permis de concevoir un premier prototype. Nous avons ensuite travaillé sur une représentation OWL pour ses propriétés d'interopérabilité. Ce travail doit encore être approfondi. Le but n'est pas de représenter automatiquement le thésaurus en OWL mais de représenter le thésaurus dans un langage comme OWL. Ainsi, les travaux décrits dans [11] et [12] ont abordé ce thème de recherche lié à cette transformation. Plus récemment, [13] en transformant le thésaurus NCI en OWL DL ont rencontré des problèmes de représentation de connaissances dont nous pourrons tirer profit dans la construction de l'ontologie.

Nous allons aborder dans la partie suivante la démarche adoptée pour construire le thésaurus TERRIDOC puis nous nous intéresserons aux éléments de la méthodologie qui permettent de transformer le thésaurus en une ontologie.

---
[4] Nous définissons une carte de concepts comme un triplet formé de : une liste de concepts, des relations entre ces concepts et des étiquettes (optionnelles) précisant ces relations.

# 4 Du fonds documentaire indexé à l'ontologie

Nous présentons à travers un exemple la méthodologie adoptée pour enrichir un premier vocabulaire de termes provenant du thésaurus RAMEAU afin de créer un thésaurus adapté, et les étapes à suivre pour transformer ce thésaurus en une ontologie.

## 4.1 Représentation sémantique de connaissances expertes

Nous nous appuyons dans notre démarche sur la base de notices descriptives correspondantes aux documents (figure 1) ainsi que sur le thésaurus RAMEAU. Dans notre phase d'extraction et de structuration des connaissances, l'exploitation des relations va nous permettre de construire le thésaurus TERRIDOC.
La première étape du traitement consiste à identifier et extraire automatiquement tous les termes (autorités matières RAMEAU) utilisés pour décrire le contenu du document dans les notices descriptives XML. Lors de la phase d'indexation, ces autorités sont sélectionnées par les bibliothécaires dans RAMEAU et utilisées dans les notices via la ou les balise(s) DEE (figure 1).

```
<DEE>Stations climatiques, thermales, etc. -- Barèges (Hautes-Pyrénées) -- 18e siècle</DEE>
<DEE>Eaux minérales -- Pyrénées (France) -- 18e siècle</DEE>
<TITRE>Précis d'observation sur les eaux de Barèges et les eaux minérales de Bigorre et du
Béarn</TITRE>
<LEGENDE> Théophile de Bourdeu est à l'origine de la mode du thermalisme pyrénéen</LEGENDE>
```

Figure. 1 – Extrait de notice descriptive 1

Chaque balise DEE correspond à une vedette-matière composée d'une ou plusieurs autorités séparées par l'élément « -- ». Chaque vedette-matière correspond à un thème estimé par l'expert comme important (l'autorité décrivant le thème est utilisée en tête de vedette) pour la description du contenu du document (*Stations climatiques, thermales, etc.* et *Eaux minérales* dans la figure 1). Nous obtenons en résultat de ce premier traitement un ensemble de termes. L'extraction automatique de cet ensemble de termes et leur mise en correspondance grâce au thésaurus RAMEAU dans un graphe conceptuel nous permet d'obtenir une première représentation sémantique du fonds documentaire. En exploitant le thésaurus RAMEAU, nous enrichissons automatiquement le vocabulaire obtenu ci-dessus avec : (i) les termes « génériques » et « employés pour » ; (ii) les relations qui leurs sont associées ; (iii) les relations entre termes associés s'il en existe.
Il faut noter que les relations hiérarchiques incluent la relation générique (genre-espèce), la relation partitive (tout-partie), la relation d'instance et les relations poly-hiérarchiques. Les travaux de D.H. Fischer [15] soulignent cette ambiguïté par le fait que la définition de ces relations « terme plus spécifique », « terme plus générique » est orientée par l'utilisation faite des thésaurus, c'est-à-dire l'aide au travail du documentaliste (indexation, recherche), et non par la formalisation de la connaissance du domaine. Nous ne pourrons lever toutes les ambiguïtés liées aux relations « terme plus spécifique », « terme plus générique ». Si nous prenons l'exemple de la relation *Lieu de villégiature >GENERIQUE> Tourisme* , elle peut sembler incorrecte. Le but premier est d'avoir un outil de navigation à travers toute la structure sémantique.
Nos premières expérimentations ont été menées sur un corpus de 750 notices descriptives et leurs documents associés relatant du patrimoine culturel pyrénéen. Nous obtenons un ensemble de 1449 termes que nous enrichissons ensuite par les termes « employés pour », « associés » et « génériques » joints et par les relations correspondantes en nous appuyant sur RAMEAU. Le thésaurus obtenu offre une première structure synthétique représentant le travail des bibliothécaires. Seule, cette structure ne peut être exploitée par les experts pour observer et analyser l'ensemble des saisies. Nous en proposons donc une représentation sous forme de carte de concepts (figure 2) pour permettre aux experts d'appréhender de façon synthétique l'ensemble du travail d'indexation d'un fonds documentaire donné, réalisé par les différents bibliothécaires y ayant contribué.

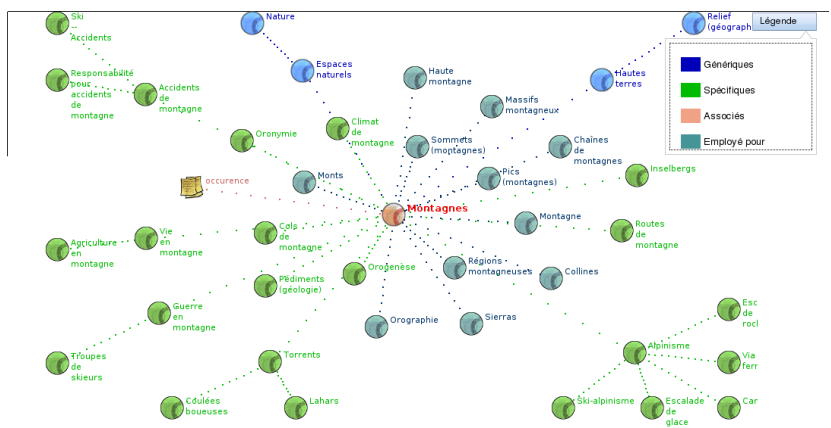

Figure. 2 – Extrait du thésaurus TERRIDOC

## 4.2 Transformation du thésaurus en ontologie légère

Une contribution importante de notre travail concerne les étapes permettant de déceler puis de labéliser les ES ainsi que les relations associatives entre ces dernières et les concepts identifiés à partir du thésaurus RAMEAU. Dans la volonté de définir une représentation d'un territoire en traitant la composante spatiale, la première étape de notre traitement consiste à identifier dans le thésaurus TERRIDOC les termes correspondants à des ES qui vont devenir des instances, et les autres qui deviendront concepts. Nous utilisons pour cela la base Système d'Informations Géographiques de l'IGN, contenant la majorité des entités nommées spatiales françaises. Par exemple, les termes du thésaurus TERRIDOC «Bagnères-de-Bigorre (Hautes-Pyrénées)» et «Barèges (Hautes-Pyrénées) » sont identifiés comme entités spatiales, ce qui nous permet de créer un concept «entité spatiale» ainsi qu'une relation d'instance *instance_of* entre le concept entité spatiale et les deux instances «Bagnères-de-Bigorre (Hautes-Pyrénées)» et «Barèges (Hautes-Pyrénées)». Les autres termes du thésaurus sont ensuite définis comme concepts en leur ajoutant comme propriétés les définitions et/ou explications provenant du thésaurus RAMEAU. Dans l'extrait de l'ontologie (figure 4), «eaux minérales» et «stations climatiques, thermales, etc.» sont ainsi définis en tant que concepts. Cela nous permet de préciser les relations génériques avec les instances «Bagnères-de-Bigorre (Hautes-Pyrénées)» et «Barèges (Hautes-Pyrénées)» en relations d'instance *instance_of*. Cette première règle nous permet, en nous appuyant sur une ressource externe type SIG, de définir une ontologie légère offrant une première représentation sémantique d'un territoire. De ce fait, l'ontologie créée permet de faire les inférences élémentaires découlant de la taxonomie des concepts (p.ex. l'héritage des propriétés) sur ces concepts particuliers.

En plus du travail de description du contenu des documents, les notices descriptives renferment des informations riches pouvant décrire un territoire. Nous proposons en deuxième étape une chaîne de traitement linguistique (syntaxique et grammaticale) afin de capturer les ES ainsi que tous les termes les qualifiant. Afin de repérer des relations sémantiques [16], nous utilisons des patrons lexico-syntaxiques. Un patron lexico-syntaxique représente une expression régulière, formée de mots, de catégories grammaticales ou sémantiques, et de symboles. Il permet d'extraire des éléments de texte respectant l'expression. Dans notre cas, les patrons exploitent les étiquettes morpho-syntaxiques ou sémantiques attribuées par Linguastream (figure 3).

*Figure 3. extrait du traitement linguistique*

La dernière phase de l'approche consiste à associer ces termes identifiés à l'ontologie par des relations de sens contenues dans les notices descriptives. Ainsi en reprenant les extraits de notices présentées figure 1, sont aussi retenus comme entités spatiales candidates les entités nommées «Bigorre» et «Béarn» que nous validons ensuite en tant qu'ES via l'appel au SIG. Un lien sémantique est alors créé entre le concept «Eaux minérales» et les instances de type spatial «Bigorre» et «Béarn» que l'on nomme *instance_of* (Figure 4).

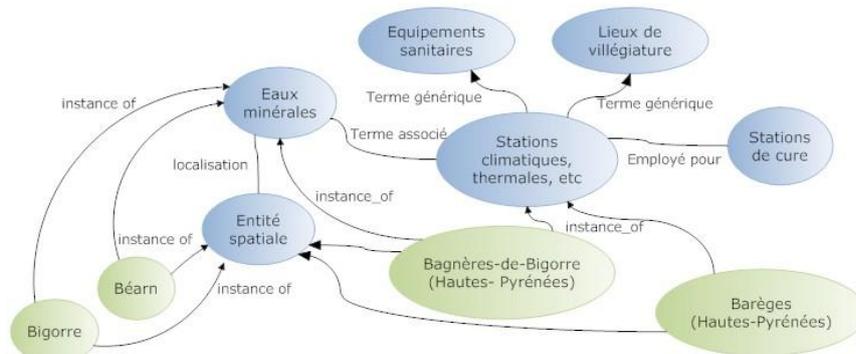

*Figure 4. Extrait de l'ontologie générée*

Nos travaux actuels cherchent à typer explicitement dans l'ontologie les relations classiques provenant du thésaurus TERRIDOC. Dans notre cas, un SIG peut nous permettre d'identifier, par calculs topologiques et géométriques sur les instances, les relations spatiales entre concepts. Nous cherchons aussi à caractériser l'ensemble des termes RAMEAU qui ne sont pas identifiés comme des instances de type spatial sous forme de concepts (possédant un nom, des caractéristiques propres sous forme d'attributs, etc.).

# 5 Conclusion

Comme nous l'avons expliqué, notre premier objectif est d'expliciter la sémantique informelle du thésaurus autour de concepts décrivant un territoire en se restreignant à l'aspect spatial dans le but de spécifier des relations ambigües entre termes présentes dans les thésaurus. Les traitements effectués sur le corpus de documents mis à disposition par la MIDR nous ont permis de modéliser la phase de création d'une ontologie enrichie du territoire reposant sur quatre étapes principales : (i) l'extraction d'informations du corpus via les notices XML associées aux documents que l'on organise sous forme d'un vocabulaire contrôlé, (ii) la définition d'un thésaurus (thésaurus TERRIDOC) caractérisant le territoire issu du vocabulaire contrôlé et du thésaurus RAMEAU (thésaurus qui a assisté l'indexation manuelle par des documentalistes des documents du fonds documentaire de la MIDR), (iii) l'identification de toponymes du territoire et de relations les associant aux concepts déjà présents dans le thésaurus, (iv) la transformation du thésaurus en ontologie légère en l'enrichissant par les toponymes et relations spatiales identifiées.

Le processus mis en place est automatique. Notre ontologie est utilisée à des fins de recherche d'information et d'explicitation du travail des experts bibliothécaires. Aussi, nous travaillons actuellement à la spécification des relations que nous créons lors de la phase de transformation d'un thésaurus en ontologie. Nous envisageons d'appliquer notre méthodologie sur le thésaurus RVM Laval afin de conforter notre approche.

# Bibliographie